
\magnification = 1200
\tolerance 1000
\pretolerance 1000
\baselineskip 20pt
\overfullrule = 0pt
\hsize=16.3truecm
\def\sp{\kern 1em}

\centerline {\bf QUANTUM RESONANCES OF WEAKLY LINKED, MESOSCOPIC,  }
\centerline{\bf SUPERCONDUCTING DOTS. }

\vskip 0.5truecm

\centerline {\bf  C.J. Lambert and A. Martin, }

\centerline {\bf School of Physics and Materials}

\centerline {\bf Lancaster University}

\centerline {\bf Lancaster   LA1 4YB}

\centerline {\bf U.K.}

\vskip 1.00truecm

\noindent
ABSTRACT.

We examine quantum properties of mesoscopic,
Josephson coupled superconducting dots,
in the limit that charging effects and quantization of energy levels within
the dots are negligible, but quasi-particle transmission  into
the weak link is not. We demonstrate that quasi-particle resonances lead to
current-phase relations, which deviate markedly from those of
weak links connecting macroscopic superconductors. Results for the steady
state dc Josephson current of two
coupled dots are presented.

\vskip 3.0truecm
PACS Numbers. 72.10.Bg, 73.40.Gk, 74.50.

\vfil\eject

Recent theories of mesoscopic Josephson weak links[1,2] have revealed a variety
of new quantization phenomena, which are absent from their macroscopic
counterparts. For such systems, the size of the junction is less than or of
order the phase breaking length, while the superconducting leads are
macroscopic. In this Letter, we consider the quantum properties of weak links,
formed
when the superconductors themselves are mesoscopic. Linked mesoscopic,
superconducting dots (LM dots), for which
quasi-particles as well as Cooper pairs maintain phase
coherence within the device, have been grown experimentally[3,4], although
no theory of their quantum properties currently exists.
In this Letter, we demonstrate that the current-phase relation of LM dots
is sensitive to resonant tunneling of quasi-particles into
the junction from external, current carrying leads. This leads
to a non-equilibrium distribution of quasi-particles within the weak link,
which is missing from conventional descriptions of Josephson junctions[5].
Such resonances,
which lead us to view LM dots, as electronic
analogues of Fabry-Perot interferometers, carry a current of order the
critical current and therefore significantly modify
the transport properties of such structures.

In a forthcoming publication[6] we shall
present a complete theory capable of describing the steady state properties
and slow dynamics of an arbitary number of LM dots. In this Letter we
illustrate how scattering theory can be used to extract the key physics,
by examining two such dots in one dimension,
described by the Bogoliubov - de Gennes equation
$$ {\bf H}(\underline r){\underline \Psi}( x)
= E{\underline \Psi}(x)
\sp ,\eqno{(1)}$$
\noindent
where
$$\bf{H} =
\left(\matrix{[-(\hbar^2/2m)\partial_x^2 +u(x) -\mu]
& \Delta (x) \cr
\Delta^*(x)
& -[-(\hbar^2/2m)\partial_x^2 +u(x) -\mu] \,}\right) \sp .\eqno{(2)}$$
In this equation, $\mu$ is the condensate chemical potential,
$u(x)$ the normal scattering potential and $\Delta (x)$ the superconducting
order parameter. A simple model of a pair a LM dots, shown
in figure 1, is obtained by allowing $\Delta (x)$ and $u(x)$
to be non-zero only in regions of size $L_1$ and $L_2$, where the
order parameter phase takes values $\phi_1$ and $\phi_2$ respectively.
Figure 1 shows a pair of LM dots connected by perfect, normal leads,
to external reservoirs at chemical potentials $\mu_a$ and $\mu_b$ and
distinguishes the present approach from
other descriptions[1,2,5], where
the sources of charge are of no consequence and the lengths $L_1$ and $L_2$
are taken to be infinite.  For LM dots, the system size $L_1 +L + L_2$ is
assumed to be smaller than the quasi-particle phase breaking length and
therefore a description, which incorporates quasi-particle phase coherence
throughout the device is appropriate.

To obtain such a description, consider the most general eigenstate of ${\bf H}$
belonging to eigen-energy $E$. In the regions where $\Delta (x)$ and
$u(x)$ vanish, this has the form

$${\underline \Psi}(x)=\cases{
\left(\matrix{A{\rm exp}[i k x] + B{\rm exp}[-i k x]\cr
C{\rm exp}[i q x] + D{\rm exp}[-i q x]}\right),&for $x<-(L_1+L/2)$;\cr
\cr
\left(\matrix{A^{''}{\rm exp}[i k x] + B^{''}{\rm exp}[-i k x]\cr
C^{''}{\rm exp}[i q x] + D^{''}{\rm exp}[-i q x]}\right),&for$-L/2<x<L/2$; \cr
\cr
\left(\matrix{A'{\rm exp}[i k x] + B'{\rm exp}[-i k x]\cr
C'{\rm exp}[i q x] + D'{\rm exp}[-i q x]}\right),&for$(L_2+L/2)<x$; \cr}
\eqno{(3)} $$
where $\hbar^2k^2/2m -\mu = \mu - \hbar^2q^2/2m = E$.
In the absence of inelastic scattering, the quantum
properties of such a structure can be described in terms
of either a transfer matrix $T$ or scattering matrix $S$, defined by
$$\left(\matrix {O^\prime \cr I^\prime} \right) = T\left(\matrix {I \cr O}
\right)
\sp\sp\sp {\rm and}\sp\sp\sp
\left(\matrix{O \cr O^\prime}\right) = S\left(\matrix{I \cr I^\prime}\right)
\sp ,\eqno{(4)}$$
\noindent
where
$$\left(\matrix{O^\prime \cr I^\prime}\right) =\hbar^{1/2} \left(
\matrix{k^{1/2}A^\prime \cr q^{1/2}D^\prime \cr
k^{1/2}B^\prime \cr q^{1/2}C^\prime}\right)
\sp\sp {\rm (5a)}\sp\sp\sp
{\rm and} \sp\sp\sp
\left(\matrix{I\cr O}\right) = \hbar^{1/2}\left(\matrix{k^{1/2}A \cr q^{1/2}D
\cr k^{1/2}B \cr q^{1/2}C}\right)
\sp .\eqno{(5b)}$$
Once $T$ is known,  $S$
can be constructed and vice versa[7,8].
Both $T$ and $S$ are functionals of all physical potentials,
as well as functions of E. Since ${\bf H}$ is Hermitian, quasi-particle
probability (though not charge) is conserved, and therefore $S$ is unitary.

To describe the Josephson effect for such a structure, one must compute the
current in the region $-L/2 < x < L/2$, as well as in the external leads.
To this end we form the expectation
value of the current density operator, with
respect to the  density matrix
corresponding to incident distributions
$f^\alpha_i(E)$ of quasi-particles of type $\alpha$, energy $E$ along lead
$i$, where $\alpha = +1 (-1)$ for particles (holes). In the simplest case,
$f^\alpha_i(E)$ would be a Fermi distribution, although the analysis outlined
here can equally well describe transport properties arising from
non-equilibrium (eg. hot electron) reservoir distributions. Expressions for the
currents $I_l$, $I_r$ in the left and right leads respectively are
written down in reference [9].
To obtain  the current within the weak link, we introduce
separate transfer matrices $T_1$, $T_2$ connecting plane wave amplitudes at
$x=0$ to those in the left and right leads respectively and satisfying
$T=T_2 T_1$. The wavefunction ${\underline \Psi}(0)$ due to, for example, an
incident particle from the left, is obtained by setting
$A=1$, $D=B'=C'=0$ in equation (3)
and using equation (4) to obtain the outgoing amplitudes $B$ and $C$. The
corresponding coefficients $A'',\sp B'',\sp C'',\sp D''$, which
we denote $A_{1+},\sp B_{1+},\sp
C_{1+},\sp D_{1+}$, respectively are then obtained by acting on the
vector (5b) with $T_1$. In this way, by combining the action of $S$ with
$T_1$ and $T_2$,  plane wave amplitudes $A_{i,\alpha},\sp B_{i,\alpha},\sp
C_{i,\alpha},\sp D_{i,\alpha}$ within the
weak link, due to a quasi-particle of type $\alpha$, incident along lead $i$
can be constructed. For simplicity we restrict the present analysis to
zero temperature, where the expectation value of
the current density operator
inside the weak link is found to be $ I_{\rm in} =  I_s + I_{qp}$,
with
$$I_s =   (2e/h)\sum_{i=1}^2\int_0^{\mu} d E \{
[\vert D_{i-}\vert^2 -
\vert C_{i-}\vert^2]+
[q( E)/k( E)][\vert D_{i+}\vert^2 -
\vert C_{i+}\vert^2]\}
\eqno{(6)}$$
and
$$\eqalign{
I_{qp}&= (2e/h)\int_0^{\mu_a-\mu} d E \{\vert A_{1+}\vert^2 -
\vert B_{1+}\vert^2 +[q( E)/k( E)][\vert C_{1+}\vert^2 -
\vert D_{1+}\vert^2]\}\cr
&+(2e/h)\int_0^{\mu-\mu_b} d E \{[k( E)/q( E)]
[\vert A_{2-}\vert^2 -
\vert B_{2-}\vert^2] +\vert C_{2-}\vert^2 -
\vert D_{2-}\vert^2\}}
\eqno{(7)}.$$
Note that the right hand side of equation (6) represents
the contribution from all occupied negative energy states, which has been
transformed, using the particle-hole symmetry relations
$ A_{i\alpha}(E) =D^*_{i-\alpha}(-E)$ and
$B_{i\alpha}(E) = -C^*_{i-\alpha}(-E)$,
to an integral over positive energies.
The division of the total current into a sum of two currents is somewhat
arbitrary. However if the reservoir potentials are equal,
$I_{qp}$ vanishes, whereas $I_s$ may remain finite. Therefore in
what follows, we refer to $I_s$ and $I_{qp}$ as the supercurrent and
quasi-particle current respectively.

Before proceeding, it worth noting that in deriving equations (6) and (7),
the occupancy of incoming states
from external reservoirs has been chosen such that
the ground state expectation value of any local operator
is preserved by the transformation from a closed to an open system.
More precisely, consider a scatter of size $L'=L_1+L+L_2$
connected to leads of size $L''$, which join together to form a closed system
of size $L'+L''$, with periodic boundary conditions. If $\rho$
is the density matrix, then the expectation value ${\rm Tr}\rho O(x)$
of an operator $O(x)$ can be  evaluated  using any
convenient set of basis states. For a closed system the obvious choice
is the set of eigenstates of ${\bf H}$, satisfying periodic boundary
conditions.
However in the limit $L'' \rightarrow \infty$, such a choice is no longer
useful and a trace over all incoming scattering states is preferred.
It is
important to note that if the incoming scattering states are interpreted
as arising from external reservoirs, then
certain properties are imposed on the  reservoirs by the scatterer.
For example at zero
temperature, in the absence of a potential difference,
expectation values are preserved  only if all incoming
quasi-particle states are populated; both incoming electron and incoming hole
states. Such occupancies are non-intuitive, since
a more natural choice of zero temperature
 reservoirs would perhaps populate only
incoming electron states. The relative merits of different choices of
incoming distributions will be discussed in a forthcoming publication[6].
Here we merely note that the existence of
 quasi-particle resonances is  independent of such choice, although the
detailed form of the current-phase relation does depend on
the distribution of incoming states.

As emphasized in reference [9], due to non-conservation
of quasi-particle charge,
the chemical potential $\mu$ enters expressions for
$I_l$ and $I_r$ explicitly and in steady state,
must be determined self-consistently
by insisting that the currents are equal. In the present context one also
notes that for an arbitrary phase difference $\phi = \phi_1 - \phi_2$,
the internal current $I_{\rm in}$ will not equal the current in the leads.
Hence to obtain a dc Josephson effect, in which a current flows between
superconducting dots of equal potentials, both $\mu$ and $\phi$ must be
determined from the steady state condition  $I_{\rm in} = I_l = I_r$.
The solution to these equations yields a phase-current expression $\phi(I)$,
which may be inverted to yield a more familiar current-phase relation
$I(\phi)$.
If the current supplied by the external
reservoirs is greater than a certain value, it may happen that no solution to
these equations exists, in which case a critical current $I_c$
has been exceeded. Since the primary aim of this Letter is to
highlight the role of quasi-particle resonances, full implementation
of this self-consistent scheme will be relegated to
reference[6]. One notes however that for
 the symmetric structures considered below,
the equation $I_l=I_r$ is trivially satisfied with the choice
$\mu = (\mu_a + \mu_b)/2$.

Except for a set of resonant energies,
the quasi-particle contribution to $I_{\rm in}$ is expected to be
negligibly small, because for reservoir potentials less than typical
values of $\vert\Delta(x)\vert$, quasi-particle states decay on the
scale of the superconducting coherence length, which in practice may
be much greater than $L_1$ and $L_2$. For a clean junction,
in the limit $L_1, L_2 \rightarrow
\infty$, these resonances correspond to bound state energies
of the weak link considered by Bardeen and Johnson[10]. In the latter
description, the penetration of quasi-particles into
the weak link is ignored and therefore in the clean limit,
the analysis presented here
reduces to that of reference[10] in the absence of quasi-particle transmission
through the superconductors.

When solving equation (1), it is convenient to
introduce a characteristic wavevector $k_F$ through the relation
$\hbar^2 k_F^2/2m = \mu$ and to divide both sides by $\mu$. The resulting
equation depends only on the dimensionless quantities $\bar E= E/\mu$,
$\bar u(x) = u(x)/\mu$, $\bar\Delta(x) = \Delta(x)/\mu$ and
$\bar x = xk_F$. With this choice of
scaling, equation (1) and therefore all scattering properties, do
not depend explicitly on $\mu$ and it is convenient to express results
in terms of the
dimensionless potential differences $\bar \mu_a= (\mu_a - \mu)/\mu$,
$\bar \mu_b= (\mu - \mu_b)/\mu$.
To illustrate the effect of resonances on the current-phase relationship
of LM dots, figure 2 shows plots of
$I_{\rm in}(\phi)$ versus $\phi$ for different choices
of $\mu$.
Choosing $\vert\bar\Delta(x)\vert =\bar\Delta_0$
in the superconducting regions of dimensionless length $\bar L_1, \bar L_2$,
figure 2 shows results for the case $\bar L =100$, $\bar L_1=\bar L_2=150$,
$\bar\Delta_0=0.04$, $\bar u(x)=0$ and an applied potential of
$\bar\mu_{ab}= (\mu_a - \mu_b)/\mu = 0.01$.
These values are typical of those attainable experimentally,
where $\bar\Delta_0$ is always much smaller than unity and the
size of a dot is much greater than the
coherence length $\bar\xi = \bar\Delta_0^{-1}$.
The dotted, dashed and solid lines
show results for $\bar\mu_b = 0, \bar\mu_{ab}/4$ and $\bar\mu_{ab}/2$
respectively. The main part of the figure shows
the internal current $I_{\rm in}(\phi) = I_s(\phi) + I_{qp}(\phi)$,
while the insert shows the quasi-particle current.
Except near the endpoints, the supercurrent $I_s$ increases
almost linearly with phase, as $\phi$ increases from $-\pi$ to $\pi$,
reflecting the fact that under certain conditions,
a phase difference across such a junction
acts like a Galilean transformation[10]. In contrast $I_{qp}$ exhibits
strong resonances over well defined intervals of  $\phi$. In the
limit $\bar L_1, \bar L_2 \rightarrow \infty$, when $\bar L >>
\bar\Delta_0^{-1}$,
the bound state energies of such a junction occur at[10]
$\bar E_n \simeq \pm\phi/\bar L +(2n+1)\pi/\bar L $.
As noted earlier these bound state energies become quasi-particle
transmission resonances in a LM dots. For $\phi = \pm\pi$, such a resonance
occurs at $E=0$ and therefore quasi-particles from external
reservoirs can enter the weak link. To understand these  results in more
detail, consider first the quasi-particle current shown by the dashed line,
arising when $\bar\mu_b = \bar\mu_{ab}/4$ and therefore $\bar\mu_a =
3\bar\mu_{ab}/4$.
As $\phi$ increases from $\phi =-\pi$, the lowest resonant
energy $\bar E_0$ first exceeds the highest incident quasi-particle
energy $\bar\mu_b$ from the right reservoir, leading to a decrease in the
quasi-particle current by a factor of $1/2$. As $\phi$ increase further,
eventually the highest energy
$\bar\mu_a$ of a quasi-particle from the left reservoir is exceeded
 and $I_{qp}$  switches off. For the case $\bar\mu_a = \bar\mu_b$ shown
 by the solid line, the quasi-particle currents from the two reservoirs switch
 off at the same phase, while for the case $\bar\mu_b=0$, only the left
 reservoir contributes to $I_{\rm qp}$.

It should be noted that the steps in $I_{\rm qp}$ are of order the critical
current through the device, despite the fact that the superconductors are
several coherence lengths long. This arises because resonant states within
the weak link are formed from superpositions of particles and Andreev
reflected holes, for which the currents add constructively. In a forthcoming
publication, it will be shown that adding normal potential scatterers to the
contacts
between the external leads and superconductors does not significantly affect
Andreev scattering within the weak link
and therefore has only a marginal effect on the quasi-particle current,
whereas potential scattering within the weak link suppresses both the
supercurrent and the quasi-particle
current.

Figure 2 illustrates the effect of quasi-particle resonances for one
value of the external potential difference only.
For
larger values of
$\mu_a - \mu_b$, more than one resonance can contribute and as shown by the
solid lines of figure
3, the resulting internal current exhibits a non-trivial phase
dependence. For completeness, the dashed lines show
the external current in the leads and from the fact that these
lines cross at several distinct values of $\phi$, one concludes
that the fully  self-consistent solution will possess many branches.

In this Letter, we have shown how current phase relations for LM dots can
be obtained from a knowledge of the transfer matrices $T_1$, $T_2$ and the
associated scattering matrix $S$. The results obtained highlight the role of
quasi-particle resonances in determining junction properties.
Figures 2 and 3 demonstrate that the quantum properties of
LM dots can  differ markedly from those of more conventional Josephson
junctions, a feature which should manifest itself in a range of
junction properties.
A key property of the current-phase relations obtained in this
Letter is that $I_{\rm in}(\phi)$ is no longer an
odd function of $\phi$ and therefore its the phase average does not vanish.
While one might be tempted to identify this phase average with a quasi-particle
leakage current, it should be emphasized that it
cannot be replaced by a phenomenological
Ohmic term in a RSJ equation. Indeed in situations where
the phase difference varies slowly
with time, the quasi-particle current is transmitted as a series
of pulses, reflecting the resonant nature of such electronic interferometers.

\vskip 0.5truecm
\noindent
$\underline {\hbox {\bf Acknowledgements.}}$

This work is supported by the SERC, the EC, NATO  and the MOD.
It has benefited from useful conversations with S.J. Robinson, B. Kramer,
M Dykman, J-L. Pichard, V.C. Hui, J.H. Jefferson and F. Sols. Support
from the Institute for Scientific Exchange is acknowledged.

\vfil\eject
\noindent
$\underline {\hbox {\bf Figure Captions.}}$
\vskip 0.5truecm
\noindent
{\bf Figure 1.}

The order parameter of a typical pair of LM dots, connected to external
reservoirs at chemical potentials $\mu_a$ and $\mu_b$, with
$\mu_a > \mu_b$.

\noindent
{\bf Figure 2.}

For $\bar \mu_a +\bar\mu_b = 0.01$,
this figure shows current-phase relations for a pair of LM dots of size
$\bar L_1=\bar L_2=150$, separation $\bar L=100$ order parameter
$\Delta_0=0.04$ and $u(x)=0$. The main body of the figure shows results for
the total internal current $I_{\rm in}(\phi)=I_s(\phi)+I_{ qp}(\phi)$,
arising when $\bar\mu_b = 0$ (dotted line), $\bar\mu_{ab}/4$ (dashed line)
and $\bar\mu_{ab}/2$ (solid line). The insert shows corresponding results for
the quasi-particle current $I_{ qp}(\phi)$ only.

\noindent
{\bf Figure 3.}
For the same system used in figure 2, this figure shows the internal
current $I_{\rm in}(\phi)$
(solid curves)
and the external current $I_{\rm ext}(\phi) = I_l=I_r$ (dashed curves)
for
applied potential differences of magnitude $\bar\mu_{ab}=$ 0.02
(top left), 0.04 (top right), 0.06 (lower left) and 0.08 (lower right).

\vfil\eject
\noindent
$\underline {\hbox {\bf References.}}$
\vskip 0.5truecm

\item{1.} C.W. Beenakker, Phys. Rev. Lett. {\bf 67} 3836 (1991).

\item{2.} L.I. Glazman and K.A. Matveev, Sov. Phys.-JETP Lett. {\bf 49} 659
(1989).

\item{3.} V.T. Petrashov and V.N. Antonov, Sov. Phys. JETP Lett. {\bf 54}
241, (1991)

\item{4.} V.T. Petrashov, V.N. Antonov, P. Delsing and T. Claeson,
Phys. Rev. Lett. {\bf 70} 347 (1993).

\item{5.} K.K. Likharev,  Rev. Mod. Phys. {\bf 51} 101 (1979).

\item{6.} C.J. Lambert, A. Martin and S.J. Robinson, to be published.

\item{7.} C.J. Lambert, V.C. Hui and S.J. Robinson, J.Phys.: Condens. Matter,
{\bf 5} 4187 (1993).

\item{8.} B. Kramer, Festk\"orperprobleme: Adv. Sol. St. Phys. {\bf 30} 53
(Vieweg, Braunschweig 1990).

\item{9.} C.J. Lambert, J. Phys.: Condens. Matter, {\bf 3} 6579 (1990).

\item{10.}J.  Bardeen and J.L. Johnson, Phys. Rev. {\bf B5} 72 (1972).

\end